\documentclass[onecolumn, 10pt, amsmath, amssymb, aps, prd]{revtex4-2}
\usepackage{hyperref}
\usepackage{slashed}
\usepackage{color}
\usepackage{graphicx}
\graphicspath{{figures/}}
\allowdisplaybreaks[1]
\usepackage{multirow}

\begin{document}
\title{Leptonic and semileptonic decays of mesons in the Domain model of QCD vacuum}
\author{Vladimir Voronin}
\email{voronin@theor.jinr.ru}
\affiliation{Joint Institute for Nuclear Research, 141980 Dubna, Moscow Region, Russia}
\begin{abstract}
The leptonic and semileptonic decays of mesons are investigated within the Domain model of QCD vacuum and hadronization. The Domain Model is the mean-field approach based on the statistical ensemble of almost everywhere homogeneous Abelian (anti-)self-dual gluon fields which reproduces main features of low-energy QCD and allows to deduce a nonlocal effective meson action. Using this meson action, the leptonic decay constants, form factors and branching ratios of semileptonic decays are evaluated simultaneously with masses of mesons. The results are compared to experimental data or other approaches.

\end{abstract}
\maketitle
\section{Introduction}
The leptonic and semileptonic decays are most easily accessed processes, both theoretically and experimentally, which involve quark flavor transformation due to weak interaction. This makes them ideal for extracting the magnitudes of elements of Cabibbo-Kobayashi-Maskawa (CKM) matrix $V$ from available experimental data.
While the CKM matrix $V$ concerns mixing of quarks, the experimental data are given for hadrons, which poses difficulties in extracting CKM elements, from the theoretical point of view. 
A multitude of methods can be employed in order to address this problem (see e.g.~\cite{Hill:2006ub,FlavourLatticeAveragingGroupFLAG:2021npn,Richman:1995wm,HFLAV:2022esi,Ivanov:1998ms} and references therein). Among them are Lattice QCD, heavy quark effective theory, Dyson-Schwinger equations, sum rules and various quark models.

In this work, leptonic and semileptonic decays of mesons are investigated within the Domain Model of QCD vacuum and hadronization~\cite{Efimov:1995uz,Burdanov:1996uw,Kalloniatis:2003sa,Nedelko:2016gdk,Nedelko:2016vpj,Nedelko:2014sla,Nedelko:2020bba} which describes the composite nature of mesons with nonlocal meson-quark interaction. 
The Domain Model of QCD vacuum and hadronization consistently describes main features of low-energy QCD.
The translation-invariant parts of gluon and quark propagators are entire analytical functions of complex momentum which can be interpreted as confinement of dynamical quarks~\cite{Efimov:1995uz}. It was shown in Ref.~\cite{Kalloniatis:2001dw} that the vacuum ensemble also provides the area law for the Wilson loop, that is the confinement of static quarks. The vacuum also provides chiral symmetry breaking and resolution of $U_A(1)$ problem~\cite{Kalloniatis:2003sa}. The mean-field model of hadronization in the presence of Abelian (anti-)self-dual vacuum gluon fields developed in Refs.~\cite{Efimov:1995uz,Burdanov:1996uw,Nedelko:2016gdk} allows to deduce an effective meson action via hadronization of one-gluon exchange of quark currents. The resulting collective colorless excitations describe extended (non-pointlike) mesons. It was shown that masses of mesons in the model exhibit Regge character at large orbital and radial quantum numbers~\cite{Efimov:1995uz}. The model describes masses of light, heavy-light mesons and heavy quarkonia~\cite{Burdanov:1996uw}, leptonic decay constants of pseudoscalar mesons and electromagnetic transition constants of vector mesons~\cite{Nedelko:2016gdk}, decay constants of vector mesons into a couple of pseudoscalar ones~\cite{Nedelko:2016vpj}, electromagnetic transition form factors of pseudoscalar mesons~\cite{Nedelko:2016vpj}, dipole polarizabilities of pseudoscalar mesons~\cite{Nedelko:2022jso}. The model was also applied to the anomalous magnetic moment of muon, in particular to dominating contributions due to strong interactions~\cite{Nedelko:2021dsh}. The present work adds leptonic decay constants of vector mesons and semileptonic form factors to the list of phenomena investigated with the Domain Model of QCD vacuum and hadronization. See also Refs.~\cite{Ivanov:2019nqd,Dubnicka:2023osn} summarizing results concerning weak interactions of mesons which were obtained within a related nonlocal model.

The paper is organized as follows.
Section~\ref{section_description_of_the_model} contains description of the model.
The leptonic decays of mesons are considered in Section~\ref{section_leptonic_decays}, and semileptonic decays in Section~\ref{section_semileptonic_decays}. The results are summarized in Section~\ref{section_summary}.

\section{Description of the Domain Model\label{section_description_of_the_model}}

The effective meson action of the Domain Model of QCD vacuum and hadronization~\cite{Efimov:1995uz,Burdanov:1996uw,Kalloniatis:2003sa,Nedelko:2016gdk,Nedelko:2016vpj,Nedelko:2014sla,Nedelko:2020bba} is given by the following formulas in Euclidean space-time:
\begin{align}
\label{meson_pf}
Z&={\cal N}
\int D\phi_{\cal Q}
\exp\left\{-\frac{\Lambda^2}{2}\frac{h^2_{\cal Q}}{g^2 C^2_\mathcal{Q}}\int d^4x 
\phi^2_{\cal Q}(x)
-\sum\limits_{k=2}^\infty\frac{1}{k}W_k[\phi]\right\},
\\
\label{effective_meson_action}
W_k[\phi]&=
\sum\limits_{{\cal Q}_1\dots{\cal Q}_k}h_{{\cal Q}_1}\dots h_{{\cal Q}_k}
\int d^4x_1\dots\int d^4x_k
\Phi_{{\cal Q}_1}(x_1)\dots \Phi_{{\cal Q}_k}(x_k)
\Gamma^{(k)}_{{\cal Q}_1\dots{\cal Q}_k}(x_1,\dots,x_k),
\\
\nonumber
\label{physical_meson_fields}
\Phi_{{\cal Q}}(x)&=\int \frac{d^4p}{(2\pi)^4}e^{ipx}{\mathcal O}_{{\mathcal Q}{\mathcal Q}'}(p)\tilde\phi_{{\mathcal Q}'}(p),\quad C_\mathcal{Q}=C_J,\quad C^2_{S/P}=2C^2_{V/A}=\frac{1}{9}.
\end{align}
Here the index $\mathcal{Q}\equiv\{aJLn\}$ stands for all quantum number of a meson. $\Lambda$ is the strength of background gluon field related to the condensate $\langle g^2 F^2 \rangle$.  
Auxiliary fields $\Phi_{{\mathcal Q}}$ introduced during hadronization are transformed into physical meson fields $\phi_{{\mathcal Q}}$ by an orthogonal matrix ${\mathcal O}_{{\mathcal Q}{\mathcal Q}'}$, so the quadratic term $W_2[\phi]$ in Eq.~\eqref{effective_meson_action} becomes diagonal.   

Inverting the quadratic part of the effective action, one finds corresponding propagators of meson fields $\phi_\mathcal{Q}$:
\begin{equation}
\label{meson_propagator}
D_\mathcal{Q}(p^2)=h_\mathcal{Q}^{-2} \left(\frac{\Lambda^2}{g^2 C^2_\mathcal{Q}}+\tilde\Gamma^{(2)}_{\mathcal{Q}}(p^2)\right)^{-1}
\end{equation}
where $\tilde\Gamma^{(2)}_{\cal Q}$ is the two-point correlation function diagonalized with respect to all quantum numbers and $g$ is strong coupling constant. 
The constants $h_\mathcal{Q}$ defined by formula
\begin{align*}
\label{hqm}
1&=h^2_{\cal Q}
\left.\frac{d}{dp^2}\tilde\Gamma^{(2)}_{\cal Q}(p^2)\right|_{p^2=-M^2_{\cal Q}}
\end{align*}
provide that the residue at the pole of meson propagator~\eqref{meson_propagator} is equal to unity.

The general function $\Gamma^{(k)}_{{\cal Q}_1\dots{\cal Q}_k}(x_1,\dots,x_k)$ includes ``connected'' and ``disconnected'' terms correlated by background gluon field. For example, $\tilde\Gamma^{(2)}_{\cal QQ'}(p)$ is given by
\begin{equation}
\label{Gammak}
\Gamma^{(2)}_{{\cal Q}_1{\cal Q}_2}=
\overline{G^{(2)}_{{\cal Q}_1{\cal Q}_2}(x_1,x_2)}-
\Xi_2(x_1-x_2)\overline{G^{(1)}_{{\cal Q}_1}G^{(1)}_{{\cal Q}_2}},
\end{equation}
Here $\Xi$ is correlation function of the background field which characterizes the statistical ensemble of the almost everywhere homogeneous Abelian (anti-)self-dual fields.
$G^{(k)}_{{\cal Q}_1\dots{\cal Q}_k}$ are quark loops averaged over background field with measure  $d\sigma_B$:
\begin{eqnarray}
\label{barG}
&&\overline{G^{(k)}_{{\cal Q}_1\dots{\cal Q}_k}(x_1,\dots,x_k)}
=\int d \sigma_B
{\rm Tr}V_{{\cal Q}_1}\left(x_1\right)S\left(x_1,x_2\right)\dots
V_{{\cal Q}_k}\left(x_k\right)S\left(x_k,x_1\right),
\\
&&\overline{G^{(l)}_{{\cal Q}_1\dots{\cal Q}_l}(x_1,\dots,x_l)
G^{(k)}_{{\cal Q}_{l+1}\dots{\cal Q}_{l+k}}(x_{l+1},\dots,x_{l+k})}
=
\nonumber\\
\nonumber
&&\int d \sigma_B
{\rm Tr}\left\{
V_{{\cal Q}_1}\left(x_1\right)S\left(x_1,x_2\right)\dots
V_{{\cal Q}_k}\left(x_l\right)S\left(x_l,x_1\right)
\right\}\times
\\
&&
{\rm Tr}\left\{
V_{{\cal Q}_{l+1}}\left(x_{l+1}\right)S\left(x_{l+1},x_{l+2}\right)\dots
V_{{\cal Q}_{l+k}}\left(x_{l+k}\right)S\left(x_{l+k},x_{l+1}\right)
\right\}.
\nonumber %\label{barGG}
\end{eqnarray}
Here $S(x,y)$ is a quark propagator in background gluon field, and $V_{{\cal Q}}$ is a nonlocal meson-quark vertex.

It is possible to find analytical expressions for propagators and vertices if one approximates the ensemble of of almost everywhere homogeneous (anti-)self-dual Abelian background gluon field with just homogeneous field. The averaging over the ensemble is then implemented by averaging the quark loops~\eqref{barG} over configurations of homogeneous gluon field. These include self-dual and anti-self-dual fields with different directions in Euclidean and color spaces. The averaging over spatial directions can be found with the help of generating formula
\begin{equation}
\label{averaging_over_vacuum_field}
\int d\sigma_B\exp(if_{\mu\nu}J_{\mu\nu})=\frac{\sin\sqrt{2\left(J_{\mu\nu}J_{\mu\nu}\pm J_{\mu\nu}\widetilde{J}_{\mu\nu}\right)}}{\sqrt{2\left(J_{\mu\nu}J_{\mu\nu}\pm J_{\mu\nu}\widetilde{J}_{\mu\nu}\right)}},
\end{equation}
where $J_{\mu\nu}$ is an arbitrary antisymmetric tensor.
Tensor $f_{\mu\nu}$ stands for an Abelian (anti-)self-dual background field with strength $\Lambda$:
\begin{gather}
\nonumber
\hat B_\mu=-\frac{1}{2}\hat n B_{\mu\nu}x_\nu, \ \hat n = t^3\cos\xi+t^8\sin\xi,
\\
\label{b-field}
\tilde{B}_{\mu\nu}=\frac12\epsilon_{\mu\nu\alpha\beta}B_{\alpha\beta}=\pm B_{\mu\nu}, \  \hat{B}_{\rho\mu}\hat{B}_{\rho\nu}=4\upsilon^2\Lambda^4\delta_{\mu\nu},\\
\nonumber
f_{\alpha\beta}=\frac{\hat{n}}{2\upsilon\Lambda^2}B_{\alpha\beta}, \  \upsilon=\mathrm{diag}\left(\frac16,\frac16,\frac13\right), \ f_{\mu\alpha}f_{\nu\alpha}=\delta_{\mu\nu}.
\end{gather}
The upper sign in ``$\pm$'' should be taken for self-dual field, and the lower for anti-self-dual field.

Nonlocal vertices $V^{aJln}_{\mu_1\dots\mu_l}$ are given by the following formulas:
\begin{gather}
V^{aJln}_{\mu_1\dots\mu_l}= {\cal C}_{ln}\mathcal{M}^a\Gamma^J F_{nl}\left(\frac{\stackrel{\leftrightarrow}{\cal D}^2\!\!\!
(x)}{\Lambda^2}\right)T^{(l)}_{\mu_1\dots\mu_l}\left(\frac{1}{i}\frac{\stackrel{\leftrightarrow}{\cal D}\!(x)}{\Lambda}\right),
\label{qmvert}\\
{\cal C}^2_{ln}=\frac{l+1}{2^ln!(n+l)!},\quad F_{nl}(s)=s^n\int_0^1 dt t^{n+l} \exp(st),
\nonumber\\
{\stackrel{\leftrightarrow}{\mathcal{D}}}\vphantom{D}^{ff'}_{\mu}=\xi_f\stackrel{\leftarrow}{\mathcal{D}}_{\mu}-\ \xi_{f'}\stackrel{\rightarrow}{\mathcal{D}}_{\mu}, 
\ \
\stackrel{\leftarrow}{\mathcal{D}}_{\mu}\hspace*{-0.3em}(x)=\stackrel{\leftarrow}{\partial}_\mu+\ i\hat B_\mu(x),  \ \ 
\stackrel{\rightarrow}{\mathcal{D}}_{\mu}\hspace*{-0.3em}(x)=\stackrel{\rightarrow}{\partial}_\mu-\ i\hat B_\mu(x), 
\nonumber\\
\xi_f=\frac{m_{f'}}{m_f+m_{f'}},\ \xi_{f'}=\frac{m_{f}}{m_f+m_{f'}}.
\nonumber
\end{gather}
Here $\mathcal{M}^a$ is a flavor matrix for a given meson, $\Gamma^J$ is corresponding Dirac matrix
\begin{equation*}
\Gamma^S=1,\ \Gamma^P=i\gamma_5,\ \Gamma^V_\mu=\gamma_\mu,\ \Gamma^A_\mu=\gamma_5\gamma_\mu,
\end{equation*}
$\xi_f,\xi_{f'}$ provide that $x$ is a center of mass for quarks with flavors $f$ and $f'$, and $n,l$ are radial and orbital quantum numbers, correspondingly. Function $F_{nl}$ is defined by the propagator of gluons charged with respect to the background field, $T^{(l)}$ are irreducible tensors of four-dimensional rotation group.
Propagator of the quark with mass $m_f$ in the presence of the homogeneous Abelian (anti-)self-dual gluon field  has the form 
\begin{align}
\label{quark_propagator}
S_f(x,y)&=\exp\left(-\frac{i}{2}\hat n x_\mu  B_{\mu\nu}y_\nu\right)H_f(x-y),
\\
\tilde H_f(p)&=\frac{1}{2\upsilon \Lambda^2} \int_0^1 ds e^{(-p^2/2\upsilon \Lambda^2)s}\left(\frac{1-s}{1+s}\right)^{m_f^2/4\upsilon \Lambda^2}
\nonumber\\
&\quad\times \left[\vphantom{\frac{s}{1-s^2}}p_\alpha\gamma_\alpha\pm is\gamma_5\gamma_\alpha f_{\alpha\beta} p_\beta
+m_f\left(P_\pm+P_\mp\frac{1+s^2}{1-s^2}-\frac{i}{2}\gamma_\alpha f_{\alpha\beta}\gamma_\beta\frac{s}{1-s^2}\right)\right],
\nonumber
\end{align}
where anti-Hermitean representation of Dirac matrices is used, and ``$\pm$'' is the same as in formula~\eqref{b-field}, $P_\pm=(1\pm\gamma_5)/2$ is the chirality projector. The translation-invariant part $H_f$ of the propagator is an analytical function in finite momentum plane which is interpreted as confinement.

\begin{figure}
\includegraphics[scale=0.25]{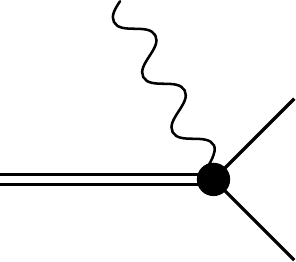}\hspace*{1em}
\includegraphics[scale=0.25]{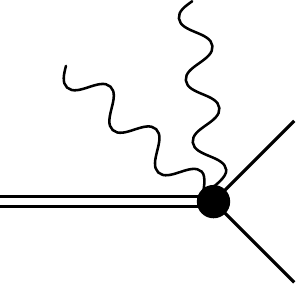}\hspace*{1em}
$\cdots$
\caption{Nonlocal meson-quark interaction with gauge bosons. The vertex with one gauge boson is of the first order in the gauge coupling constant, the vertex with two bosons is of the second order, and so on.\label{figure_nonlocal}}
\end{figure}

A common way of extracting matrix $V$ from experimental data is via matrix elements of quark currents. For example, leptonic decay constants of mesons are found from matrix element
\begin{equation}
\label{leptonic_current_element}
\langle 0| \bar{q}'\gamma_\mu \gamma_5 q|H\rangle.
\end{equation}
The formula~\eqref{leptonic_current_element} is known as impulse approximation which is often taken as a definition for corresponding amplitudes (see e.g. review~\cite{FlavourLatticeAveragingGroupFLAG:2021npn}). 
It is known that besides an ordinary boson-current interaction, in bound-state problems there is also an additional interaction with a gauge field which appears when the gauge invariance and current conservation are introduced in a consistent way~\cite{Terning:1991yt,Gross:1987bu,Woloshyn:1975vg,Bentz:1985uji}. The diagrammatical representation of these additional terms is shown in Fig.~\ref{figure_nonlocal}.
When one includes photons and weak gauge bosons into the Domain Model, the interactions shown in Fig.~\ref{figure_nonlocal} emerge, and their contributions to the amplitudes can be explicitly calculated, as well as those given by the impulse approximation.

Electromagnetic and weak interactions are introduced into the meson action in Eq.~\eqref{effective_meson_action} in a gauge-invariant way with the prescription outlined in Ref.~\cite{Terning:1991yt}. After this procedure and hadronization, the meson action includes local interactions contained in the Lagrangian of the Standard Model and additional terms due to nonlocality (see Refs.~\cite{Nedelko:2016gdk,Nedelko:2021dsh}). The term relevant for the present paper yields a nonlocal meson-quark vertex with one charged gauge boson $W^+$,
\begin{multline}
\label{qmvert_W}
V^{aJln;\mu}_{f_1 f_2}(x;q)= -\frac{g}{\sqrt{2}} \int_0^1 d\tau \frac{1}{\tau} \frac{\partial}{\partial q_\mu}\\
\left\{P_R \left(
\begin{array}{cc}
0 & V^\text{CKM}\\
0 & 0
\end{array}
\right)_{f_1 f} V^{aJln}_{ff'}\left( \stackrel{\leftrightarrow}{\mathcal{D}}(x) -iq\tau\xi\right) \delta_{f'f_2} -
\delta_{f_1f} V^{aJln}_{ff'}\left( \stackrel{\leftrightarrow}{\mathcal{D}}(x) + iq\tau\xi' \right) \left(
\begin{array}{cc}
0 & V^\text{CKM}\\
0 & 0
\end{array}
\right)_{f' f_2} P_L\right\},
\end{multline}
and its analog for $W^-$. Here $V^\text{CKM}$ is CKM matrix, $q$ is the momentum of $W^+$, $P_R=(1+\gamma_5)/2$ and $P_L=(1-\gamma_5)/2$ are chirality projectors, $g=e/\sin\theta_\mathrm{W}$, $e$ is electric charge and $\theta_\mathrm{W}$ is weak mixing angle. Note that vertices~\eqref{qmvert_W} do not appear if one introduces weak interactions via Fermi four-fermion interaction.

Technically, the evaluation of one-loop diagrams reduces to Gaussian integrals over coordinates or momenta, averaging over background field using Eq.~\eqref{averaging_over_vacuum_field}, analytical continuation to Minkowski space-time, and numerical integration over remaining proper times which appear in Eqs.~\eqref{qmvert},~\eqref{quark_propagator},~\eqref{qmvert_W}. The representation of vertex operator $F_{nl}$ given in the Appendix helps reduce the complexity of Gaussian integrations within the computer algebra systems such as FORM~\cite{Kuipers:2012rf} which was extensively used for evaluation of the amplitudes investigated in the present work.

The masses of mesons can be found from poles of corresponding propagators~\eqref{meson_pf}. The masses of $\pi,\rho,K,K^*,D,B$ are used to extract the parameters of the model (see Ref.~\cite{Nedelko:2016gdk} for details): scale $\Lambda$ related to gluon condensate $\langle g^2F^2\rangle$, quark masses and strong constant $\alpha_s$.
\begin{table}
\begin{center}
\begin{tabular}{|@{\hspace*{1em}\extracolsep{0.5em}}cccccc@{\hspace*{1em}\extracolsep{0em}}|}
\hline
$m_{u/d}$, MeV&$m_s$, MeV&$m_c$, MeV&$m_b$, MeV&$\Lambda$, MeV&$\alpha_s$\\
\hline
145&376&1715&5115&416&3.45\\
\hline
\end{tabular}\\
\begin{tabular}{|c|cccccc|}
\hline
&$\pi$&$K$&$\rho$&$K^*$&$D^0$&$B^0$\\
\hline
$M=M_\text{exp},\ \text{MeV}$&$139.57$&$493.67$&$775.26$&$891.66$&$1864.86$&$5279$\\
\hline
\end{tabular}\\
\begin{tabular}{|c|cccccccccc|}
\hline
&$\omega$&$\phi$&$D^*$&$D_s$&$D_s^*$&$B^*$&$B_s$&$B_s^*$&$B_c$&$B_c^*$\\
\hline
$M_\text{exp},\ \text{MeV}$&$782$&$1019.46$&$2010.28$&$1968.35$&$2112.3$&$5325$&$5366.7$&$5415.4$& $6274.47$& $6328$~\cite{Dowdall:2012ab}\\
\hline
$M,\ \text{MeV}$&$775.26$&$1039$&$2088$&$1975$&$2235$&$5452$&$5373$&$5591
$&$6312$&$6678$\\
\hline
\end{tabular}
\caption{Values of parameters fitted to masses of $\pi,\rho,K,K^*,D^0,B^0$, and masses of other mesons evaluated with these parameters and used in calculations in the paper. $M_\text{exp}$ for all mesons except $B_c^*$ are taken from Ref.~\cite{ParticleDataGroup:2022pth}
\label{parameters_of_the_model}
}
\end{center}
\end{table}
The parameters and evaluated masses are given in Table~\ref{parameters_of_the_model}. In practice, the matrix $O_\mathcal{QQ'}$ is truncated to some finite order, and in the present paper it is the matrix $O_\mathcal{QQ'}=O_{nn'}$ which mixes seven radial states. The masses of charmed and bottom quarks are extracted from $D^0$ and $B^0$ mesons instead of $J/\psi$ and $\Upsilon$ in previous papers, e.g.~\cite{Nedelko:2016gdk}. This helps reduce the errors due to phase space of semileptonic decays which depends on masses of mesons rather sharply. The leptonic decay constants of pseudoscalar mesons are recalculated in Section~\ref{section_leptonic_decays} for consistency.

The energies of decays considered below are far less than the mass of the weak gauge boson $W^\pm$, so it is sufficient to approximate the exchange of gauge boson with Fermi constant $G_F$. The values of $G_F$ and CKM matrix elements $V_{qq'}$ are taken from PDG~\cite{ParticleDataGroup:2022pth}. 

\section{Leptonic decays of mesons\label{section_leptonic_decays}}
The amplitude of leptonic decays of pseudoscalar mesons can be parametrized as
\begin{equation*}
A(H(p)\to\ell(k)\bar{\nu}_\ell(k'))=\frac{G_F}{\sqrt{2}}V_{qq'}\bar{\ell}\gamma_\mu(1-\gamma_5)\nu_\ell M_H^\mu=i\frac{G_F}{\sqrt{2}}V_{qq'}\bar{\ell}\gamma_\mu(1-\gamma_5)\nu_\ell f_Hp^\mu
\end{equation*}
and for vector mesons as
\begin{equation*}
A(H(p)\to\ell(k)\bar{\nu}_\ell(k'))=\frac{G_F}{\sqrt{2}}V_{qq'}\bar{\ell}\gamma_\mu(1-\gamma_5)\nu_\ell M_H^{\mu\nu}\vec{e}_\nu(p)=\frac{G_F}{\sqrt{2}}V_{qq'}\bar{\ell}\gamma_\mu(1-\gamma_5)\nu_\ell M f_H\vec{e}^\mu(p),
\end{equation*}
where $p^2=M^2,\quad k^2=m^2,\quad k'^2=0$.
Here $V_{qq'}$ is the element of CKM matrix corresponding to a given decay, $M$ is the mass of decaying meson, $m$ is the mass of final lepton, and $f_H$ is the constant which parametrizes hadronic part $M_H$ of the corresponding amplitude.
\begin{figure}
\includegraphics[scale=.25]{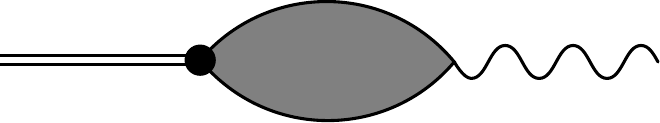}\hspace*{2em}
\includegraphics[scale=.25]{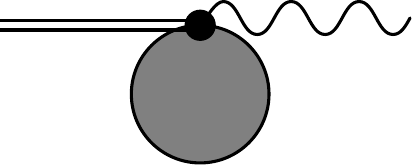}
\caption{Diagrams contributing to leptonic decay of mesons. The grey background denotes averaging over vacuum gluon field.\label{diagrams_leptonic_decay}}
\end{figure}
The diagrams contributing to the hadronic part of these decays are shown in Fig.~\ref{diagrams_leptonic_decay}. The grey background in the diagrams indicates that the background gluon field is taken into account nonperturbatively.
The contribution of these diagrams for ground-state pseudoscalar mesons is 
\begin{equation}
\label{leptonic_pseudoscalar_full}
\begin{split}
(2\pi)^4\delta^{(4)}(p-q)\frac{1}{2}V_{qq'}M_H^\mu=&\sum_n O_{n0}^{aP}\left[\int d\sigma_B \int d^4x \int d^4y\ e^{ipx-iqy}\ (-1)\mathrm{Tr}V^{aP0n}(x)S(x,y)\gamma^\mu\frac{1-\gamma_5}{2}V^\text{CKM}S(y,x)\right.\\
&\left.+\int d\sigma_B \int d^4x\ e^{ipx-iqx}\ (-1)\sum_n \mathrm{Tr}V^{aP0n;\mu}(x;-q)S(x,x)\right],
\end{split}
\end{equation}
where $q=k+k'$ is momentum of virtual $W$ boson, and $a$ is the flavor index corresponding to the meson $H$. The trace is with respect to color, flavor and spinor indices. The matrix $O_{nn'}^{aJ}$ is found from location of the pole of propagator~\eqref{meson_propagator} for each meson. Analogous expression for the vector mesons is
\begin{equation}
\label{leptonic_vector}
\begin{split}
(2\pi)^4\delta^{(4)}(p-q)\frac{1}{2}V_{qq'}M_H^{\mu\nu}=&\sum_n O_{n0}^{aV}\left[\int d\sigma_B \int d^4x \int d^4y\ e^{ipx-iqy}\ (-1)\mathrm{Tr}V_\nu^{aP0n}(x)S(x,y)\gamma^\mu\frac{1-\gamma_5}{2}V^\text{CKM}S(y,x)\right.\\
&\left.+\int d\sigma_B \int d^4x\ e^{ipx-iqx}\ (-1)\sum_n \mathrm{Tr}V_\nu^{aP0n;\mu}(x;-q)S(x,x)\right],
\end{split}
\end{equation}
The first term in these expressions corresponds to the matrix element of current given by formula
\begin{equation}
\label{leptonic_current_el}
\langle 0| \bar{q}'\gamma_\mu (1-\gamma_5) q|H\rangle.
\end{equation}

It can be noted that at large Euclidean momenta $p$ the meson-quark vertex~\eqref{qmvert} behaves as $1/p^2$ if $l=0$, and the vertex with gauge boson~\eqref{qmvert_W} behaves as $1/p^3$. Therefore the diagrams in Fig.~\ref{diagrams_leptonic_decay} should logarithmically diverge. In order to regularize the divergences, one introduces a small positive shift $\varepsilon$ into the lower boundary of integration  with respect to $t$ in Eqs.~\eqref{qmvert},\eqref{qmvert_W}. Explicit calculation shows the sum of diagrams is finite after the regularization is removed, $\varepsilon\to 0$. Moreover, for pseudoscalar mesons each diagram in Fig.~\ref{diagrams_leptonic_decay} is finite on their own, so it is possible to evaluate them separately. The results of calculations for pseudoscalar mesons are given in Table~\ref{table_leptonic_pseudoscalar}, and for vector mesons in Table~\ref{table_leptonic_vector}.
\begin{table}
\begin{tabular}{|c|c|c|c|}
\hline
\multirow[c]{2}{*}{meson}&\multirow[c]{2}{*}{decay constant $f_P$, MeV}&\multicolumn{2}{c|}{$f_P$, MeV, this work}
\\
\cline{3-4}
&&impulse approximation&full\\
\hline
$\pi$&$131.7$~\cite{ParticleDataGroup:2022pth}&131.2&140.4\\
$K$&$157.3$~\cite{ParticleDataGroup:2022pth}&161.2&178.6\\
$D$&$208.5$~\cite{ParticleDataGroup:2022pth}&187.8&231.1\\
$D_s$&$251.8$~\cite{ParticleDataGroup:2022pth}&245.6&286.8\\
$B$&$205.7$~\cite{ParticleDataGroup:2022pth}&164.7&203\\
$B_s$&$230.7$~\cite{Bazavov:2017lyh}&220.7&262.8\\
$B_c$&$427\pm 6\pm 2$~\cite{McNeile:2012qf}&403.8&450.2\\
\hline
\end{tabular}
\caption{Leptonic decay constants of pseudoscalar mesons compared to available data. The term in formula~\eqref{leptonic_pseudoscalar_full} corresponding to impulse approximation and matrix element~\eqref{leptonic_current_el} is given in column ``impulse approximation'', the column ``full'' includes all contributions given in Eq.~\eqref{leptonic_pseudoscalar_full}. \label{table_leptonic_pseudoscalar}}
\end{table}
\begin{table}
\begin{tabular}{|c|c|c|}
\hline
meson&decay constant $f_V$, MeV&$f_V$, MeV, this work
\\
\hline
$\rho$&$208.5\pm 5.5\pm 0.9$~\cite{Sun:2015enu}&$225.9$\\
$\phi$&$241\pm 18$~\cite{Donald:2013pea}&$222.8$\\
$K^*$&$202.5$~\cite{ParticleDataGroup:2022pth}&$219$\\
$D^*$&$223.5\pm 8.7$~\cite{Lubicz:2017asp}&$174.3 $ \\
$D_s^*$&$268.8\pm 6.5$~\cite{Lubicz:2017asp}&$202.8$\\
$B^*$&$186.4 \pm 7.1$~\cite{Lubicz:2017asp}&$133.4$\\
$B_s^*$&$223.1\pm 5.6$~\cite{Lubicz:2017asp}&$165.8$\\
$B_c^*$&$422\pm 13$~\cite{Colquhoun:2015oha}&$299.8$\\
\hline
\end{tabular}
\caption{Leptonic decay constants of vector mesons. The calculated values are extracted from Eq.~\eqref{leptonic_vector}. \label{table_leptonic_vector}}
\end{table}

\section{Semileptonic decays of mesons\label{section_semileptonic_decays}}
The amplitude of semileptonic decay $P\to P'\ell\bar{\nu}_\ell$ with a pseudoscalar meson in the final state 
\begin{equation*}
A(H(p)\to H'(p')\ell\bar{\nu}_\ell)=\frac{G_F}{\sqrt{2}}V_{qq'}\bar{\ell}\gamma_\mu(1-\gamma_5)\nu_\ell M_{HH'}^\mu,
\end{equation*}
can be parametrized as
\begin{equation}
\label{form_factors_pseudoscalar}
M_{HH'}^\mu=F_+(q^2)P^\mu+F_-(q^2)q^\mu,
\end{equation}
where $P=p+p',q=p-p'$.
The hadronic part of the amplitude $P\to V\ell\bar{\nu}_\ell$  with a vector meson in the final state
\begin{equation*}
A(H(p)\to H'(p')\ell\bar{\nu}_\ell)=\vec{e}_\alpha^\dagger\frac{G_F}{\sqrt{2}}V_{qq'}\bar{\ell}\gamma_\mu(1-\gamma_5)\nu_\ell M_{HH'}^{\mu\alpha},
\end{equation*}
where $\vec{e}_\alpha^\dagger$ is polarization of vector meson, can be represented in the form suggested in Ref.~\cite{Wirbel:1985ji}:
\begin{gather}
\label{form_factors_vector}
\begin{split}
\vec{e}_\alpha^\dagger M_{HH'}^{\mu\alpha}=&-(M+M')\vec{e}_\alpha^\dagger g^{\mu\alpha}A_1(q^2)+ \frac{\vec{e}_\alpha^\dagger q^\alpha}{M+M'}P^\mu A_2(q^2)\\
& +2M'\frac{\vec{e}_\alpha^\dagger q^\alpha}{q^2}q^\mu \left[A_3(q^2)-A_0(q^2)\right]+\frac{2i\varepsilon^{\mu\alpha\rho\sigma}\vec{e}_\alpha^\dagger p'_{\rho} p_{\sigma}}{M+M'}V(q^2),
\end{split}\\
\nonumber
A_3(q^2)=\frac{M+M'}{2M_2}A_1(q^2)-\frac{M-M'}{2M_2}A_2(q^2).
\end{gather}
Here $A_0(0)=A_3(0)$.

\begin{figure}
\includegraphics[scale=.25]{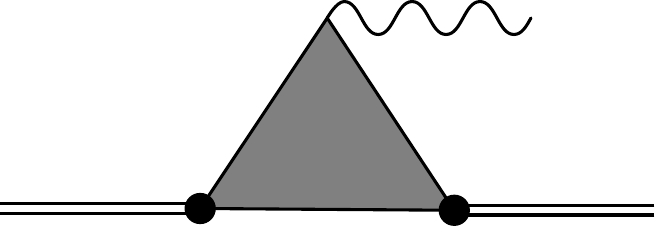}\hspace*{2em}
\includegraphics[scale=.25]{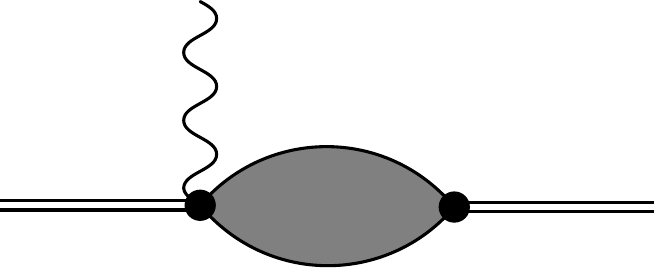}\hspace*{2em}
\includegraphics[scale=.25]{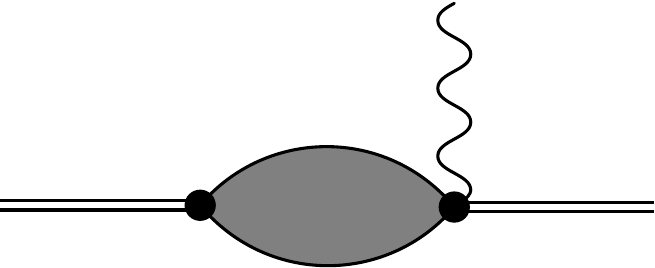}\hspace*{2em}
\caption{Diagrams contributing to semileptonic decay.\label{diagrams_semileptonic_decay}}
\end{figure}

The diagrams for semileptonic decay are shown in Fig.~\ref{diagrams_semileptonic_decay},
and their contribution is given by the formula
\begin{equation}
\label{hadronic_part_semileptonic_pseudoscalar}
\begin{split}
(2\pi)^4&\delta^{(4)}(p-p'-q)\frac{1}{2}V_{qq'}M_{HH'}^\mu=\sum_{nn'}O_{n0}^{aP}O_{n'0}^{a'P}\\
&\left[\int d\sigma_B \int d^4x \int d^4y \int d^4z\ e^{ipx-ip'y-iqz}\ (-1)\mathrm{Tr}V^{aP0n}(x)S(x,y)V^{a'P0n'}(y)S(y,z)\gamma^\mu\frac{1-\gamma_5}{2}V^\text{CKM}S(z,x)\right.\\
&+\int d\sigma_B \int d^4x \int d^4y\ e^{ipx-iqx-ip'y}\ (-1)\mathrm{Tr}V^{aP0n;\mu}(x;-q)S(x,y)V^{a'P0n'}(y)S(y,x)\\
&\left.+\int d\sigma_B \int d^4x \int d^4y\ e^{ipx-ip'x-iqy}\ (-1)\mathrm{Tr}V^{aP0n}(x)S(x,y)V^{a'P0n';\mu}(y;-q)S(y,x)\right].
\end{split}
\end{equation}
for final-state pseudoscalar mesons and
\begin{equation}
\label{hadronic_part_semileptonic_vector}
\begin{split}
(2\pi)^4&\delta^{(4)}(p-p'-q)\frac{1}{2}V_{qq'}M_{HH'}^{\mu\nu}=\sum_{nn'}O_{n0}^{aP}O_{n'0}^{a'V}\\
&\left[\int d\sigma_B \int d^4x \int d^4y \int d^4z\ e^{ipx-ip'y-iqz}\ (-1)\mathrm{Tr}V^{aP0n}(x)S(x,y)V_\nu^{a'V0n'}(y)S(y,z)\gamma^\mu\frac{1-\gamma_5}{2}V^\text{CKM}S(z,x)\right.\\
&+\int d\sigma_B \int d^4x \int d^4y\ e^{ipx-iqx-ip'y}\ (-1)\mathrm{Tr}V^{aP0n;\mu}(x;-q)S(x,y)V_\nu^{a'V0n'}(y)S(y,x)\\
&\left.+\int d\sigma_B \int d^4x \int d^4y\ e^{ipx-ip'x-iqy}\ (-1)\mathrm{Tr}V^{aP0n}(x)S(x,y)V_\nu^{a'V0n';\mu}(y;-q)S(y,x)\right].
\end{split}
\end{equation}
for vector mesons. Each term is finite, and no regularization is needed. Only the first term in Eq.~\eqref{hadronic_part_semileptonic_vector} contributes to the form factor $V$ defined by Eq.~\eqref{form_factors_vector}.
Analogously to Eqs.~\eqref{leptonic_pseudoscalar_full} and~\eqref{leptonic_vector}, the first term in the above formulas corresponds to the matrix element
\begin{equation*}
\langle H'|q'\gamma^\mu(1-\gamma_5)q|H\rangle.
\end{equation*}
Formulas~\eqref{hadronic_part_semileptonic_pseudoscalar} and~\eqref{hadronic_part_semileptonic_vector} allow to extract form factors defined with formulas~\eqref{form_factors_pseudoscalar},~\eqref{form_factors_vector}, and hence phenomenology of semileptonic decays. The differential decay rate is given by
\begin{equation*}
\frac{d\Gamma}{dq^2d\cos\theta}=\frac{p^*}{(2\pi)^3 32M^2}\left(1-\frac{m^2}{q^2}\right)\sum_\text{pol}|A|^2.
\end{equation*}
Here $p^*$ is the absolute value of final meson momentum in the rest frame of decaying meson:
\begin{equation*}
{p^*}^2=\frac{\left(M^2-\left(M'-\sqrt{q^2}\right)^2\right)\left(M^2-\left(M'+\sqrt{q^2}\right)^2\right)}{4M^2},
\end{equation*}
and $\pi-\theta$ is the angle between the momenta of final lepton and final meson in the center of mass of final lepton-neutrino pair $q=(\sqrt{q^2},0)$. 
In this reference frame integration over angle $\theta$ is trivial, and one easily finds the total semileptonic decay rate
\begin{equation}\label{semileptonic_decay_rate}
\Gamma=\int_{m^2}^{(M-M')^2}dq^2\int_{-1}^1 d\cos\theta \frac{d\Gamma}{dq^2d\cos\theta}.
\end{equation}
The results of calculations are presented in Tables~\ref{table_semileptonic_pseudoscalar} and~\ref{table_semileptonic_vector}. Branching ratios are found by dividing decay width~\eqref{semileptonic_decay_rate} by the total decay width taken from Ref.~\cite{ParticleDataGroup:2022pth}.
\begin{table}
\begin{tabular}{|c|c|c||c|c|}
\hline
\multirow[c]{2}{*}{decay}&\multicolumn{2}{c||}{available data}&\multicolumn{2}{c|}{this work}\\
\cline{2-5}
&$F_+(0)$&branching ratio&$F_+(0)$&branching ratio\\
\hline
$K^-\to \pi^0 e^- \bar{\nu}_e$&&$(5.07\pm 0.04)\times 10^{-2}$&$0.699$&$4.68\times 10^{-2}$\\
$K^-\to \pi^0 \mu^- \bar{\nu}_\mu$&&$(3.352\pm 0.033)\times 10^{-2}$&&$3.05\times 10^{-2}$\\
\hline
$K_L\to \pi^\pm e^\mp \nu_e$&$0.6856\pm 0.0010$~\cite{FermilabLattice:2018zqv}&$(4.055\pm 0.011)\times 10^{-1}$&$0.699$&$3.86\times 10^{-1}$\\
$K_L\to \pi^\pm \mu^\mp \nu_\mu$&&$(2.704\pm 0.001)\times 10^{-1}$&&$2.52\times 10^{-1}$\\
\hline
$D^0\to K^- e^+ \nu_e$&$0.7368 \pm 0.0044$~\cite{BESIII:2015tql}&$(3.549\pm 0.026)\times 10^{-2}$&$0.813$&$3.68\times10^{-2}$\\
$D^0\to K^- \mu^+ \nu_\mu$&&$(3.41\pm 0.04)\times 10^{-2}$&&$3.57\times 10^{-2}$\\
\hline
$D^0\to \pi^- e^+ \nu_e$&$0.6372 \pm 0.0091$~\cite{BESIII:2015tql}&$(2.91\pm 0.04)\times 10^{-3}$&$0.745$&$2.91\times 10^{-3}$\\
$D^0\to \pi^- \mu^+ \nu_\mu$&&$(2.67\pm 0.12)\times 10^{-3}$&&$2.85\times 10^{-3}$\\
\hline
$D^+\to \bar{K}^0 e^+ \nu_e$&$0.725 \pm 0.013$~\cite{BESIII:2017ylw}&$(8.72\pm 0.09)\times 10^{-2}$&$0.813$&$9.27\times 10^{-2}$\\
$D^+\to \bar{K}^0 \mu^+ \nu_\mu$&&$(8.76\pm 0.19)\times 10^{-2}$&&$8.99\times 10^{-2}$\\
\hline
$D^+\to \pi^0 e^+ \nu_e$&$0.440 \pm 0.009$~\cite{BESIII:2017ylw}&$(3.72\pm 0.17)\times 10^{-3}$&$0.527$&$3.66\times 10^{-3}$\\
$D^+\to \pi^0 \mu^+ \nu_\mu$&&$(3.50\pm 0.15)\times 10^{-3}$&&$3.59\times 10^{-3}$\\
\hline
$D_s^+\to K^0 e^+ \nu_e$&$0.720\pm 0.085$~\cite{BESIII:2018xre}&$(3.4\pm 0.4)\times 10^{-3}$&$0.611$&$2.18\times 10^{-3}$\\
\hline
$B^0\to D^- \ell^+ \nu_\ell$&$0.717\pm 0.05$~\cite{MILC:2015uhg}&$(2.24\pm 0.09)\times 10^{-2}$&$0.839$&$2.99\times 10^{-2}$\\
$B^0\to D^- \tau^+ \nu_\tau$&&$(1.05\pm 0.23)\times 10^{-2}$&&$7.73\times 10^{-3}$\\
\hline
$B^0\to \pi^- \ell^+ \nu_\ell$&$0.297 \pm 0.030$~\cite{Leljak:2021vte}&$(1.50\pm 0.06)\times 10^{-4}$&$0.348$&$1.83\times 10^{-4}$\\
\hline
$B^+\to \bar{D}^0\ell\bar{\nu}_\ell$&&$(2.30\pm 0.09)\times 10^{-2}$&$0.839$&$3.23\times 10^{-2}$\\
$B^+\to \bar{D}^0\tau\bar{\nu}_\tau$&&$(7.7\pm 2.5)\times 10^{-3}$&&$8.33\times 10^{-3}$\\
\hline
$B^+\to \pi^0 \ell^+ \nu_\ell$&&$(7.8\pm 0.27)\times 10^{-5}$&$0.246$&$9.85\times 10^{-5}$\\
\hline
$B_s^0\to K^- \mu^+ \nu_\mu$&$0.336 \pm 0.023$~\cite{Khodjamirian:2017fxg}&$(1.06\pm 0.09)\times 10^{-4}$&$0.270$&$1.34\times 10^{-4}$\\
\hline
$B_s^0\to D_s^-\mu^+\nu_\mu$&$0.665\pm 0.012$~\cite{McLean:2019qcx}&$(2.44\pm 0.23)\times 10^{-2}$&$0.801$&$2.87\times 10^{-2}$\\
\hline
\end{tabular}
\caption{Decays $P\to P'\ell\bar{\nu}_\ell$ or their charge conjugates. Unless indicated otherwise, the data for comparison are taken from PDG~\cite{ParticleDataGroup:2022pth}. The width of $K_L$ is the sum of the charge states. The value for $F_0^{K^-\pi^0}$ is found by dividing $F_+^{K^0\pi^-}(0)$ taken from Ref.~\cite{FermilabLattice:2018zqv} by $\sqrt{2}$.
\label{table_semileptonic_pseudoscalar}}
\end{table}
\begin{table}
\begin{tabular}{|c|c|c|c||c|c|c|}
\hline
\multirow[c]{2}{*}{decay}&\multicolumn{3}{c||}{available data}&\multicolumn{3}{c|}{this work}\\
\cline{2-7}
&$r_V$&$r_2$&branching ratio&$r_V$&$r_2$&branching ratio\\
\hline
$D^0\to \rho^- e^+ \nu_e$&$1.64\pm 0.10$&$0.84\pm 0.06$&$(1.50\pm 0.12)\times 10^{-3}$&$1.37$&$0.819$&$3.35\times 10^{-3}$\\
$D^0\to \rho^- \mu^+ \nu_\mu$&&&$(1.35\pm 0.13)\times 10^{-3}$&&&$3.20\times 10^{-3}$\\
\hline
$D^0\to K^{*-} e^+ \nu_e$&$1.46\pm 0.07$&$0.68\pm 0.06$&$(2.15\pm 0.16)\times 10^{-2}$&$1.24$&$0.788$&$4.07\times 10^{-2}$\\
$D^0\to K^{*-} \mu^+ \nu_\mu$&&&$(1.89\pm 0.24)\times 10^{-2}$&&&$3.84\times 10^{-2}$\\
\hline
$D^+\to \rho^0 e^+ \nu_e$&$1.64\pm 0.10$&$0.84\pm 0.06$&$\left(2.18\begin{array}{c}+0.17\\[-1ex]-0.25\end{array}\right)\times 10^{-3}$&$1.37$&$0.819$&$4.22\times 10^{-3}$\\
$D^+\to \rho^0 \mu^+ \nu_\mu$&&&$(2.4\pm 0.4)\times 10^{-3}$&&&$4.03\times 10^{-3}$\\
\hline
$D^+\to \omega e^+ \nu_e$&$1.24\pm 0.11$&$1.06\pm 0.16$&$(1.69\pm 0.11)\times 10^{-3}$&$1.37$&$0.819$&$4.22\times 10^{-3}$\\
$D^+\to \omega \mu^+ \nu_\mu$&&&$(1.77\pm 0.21)\times 10^{-3}$&&&$4.03\times 10^{-3}$\\
\hline
$D^+\to \bar{K}^{*0} e^+ \nu_e$&$1.49\pm 0.05$&$0.802\pm 0.021$&$(5.40\pm 0.10)\times 10^{-2}$&$1.24$&$0.788$&$1.02\times 10^{-1}$\\
$D^+\to \bar{K}^{*0} \mu^+ \nu_\mu$&&&$(5.27\pm 0.15)\times 10^{-2}$&&&$9.66\times 10^{-2}$\\
\hline
$D_s^+\to \phi e^+ \nu_e$&$1.80\pm 0.08$&$0.84\pm 0.11$&$(2.39\pm 0.16)\times 10^{-2}$&$1.36$&$0.879$&$3.54\times 10^{-2}$\\
$D_s^+\to \phi \mu^+ \nu_\mu$&&&$(1.9\pm 0.5)\times 10^{-2}$&&&$3.34\times 10^{-2}$\\
\hline
$D_s^+\to K^{*0} e^+ \nu_e$&$1.7\pm 0.4$&$0.77\pm 0.29$&$(2.15\pm 0.28)\times 10^{-3}$&$1.46$&$0.689$&$2.82\times 10^{-3}$\\
\hline
$B^0\to \rho^{-} \ell^+ \nu_\ell$&$1.270\pm 0.240$~\cite{Gubernari:2018wyi}&$0.874\pm 0.192$~\cite{Gubernari:2018wyi}&$(2.94\pm 0.21)\times 10^{-4}$&$1.17$&$0.928$&$8.14\times 10^{-4}$\\
\hline
$B^0\to D^{*-} \ell^+ \nu_\ell$&$1.151 \pm 0.114$~\cite{Gubernari:2018wyi}&$0.856 \pm 0.076$~\cite{Gubernari:2018wyi}&$(4.97\pm 0.12)\times 10^{-2}$&$1.06$&$0.926$&$6.82 \times 10^{-2}$\\
$B^0\to D^{*-} \tau^+ \nu_\tau$&&&$(1.58\pm 0.09)\times 10^{-2}$&&&$1.47\times 10^{-2}$\\
\hline
$B^+\to \rho^{0} \ell^+ \nu_\ell$&&&$(1.58\pm 0.44)\times 10^{-4}$&$1.17$&$0.928$&$4.39\times 10^{-4}$\\
\hline
$B^+\to \omega \ell^+ \nu_\ell$&$1.254\pm 0.056$~\cite{Bharucha:2015bzk}&$0.878\pm 0.081$~\cite{Bharucha:2015bzk}&$(1.19\pm 0.09)\times 10^{-4}$&$1.17$&$0.928$&$4.39\times 10^{-4}$\\
\hline
$B^+\to \bar{D}^{*0} \ell^+ \nu_\ell$&&&$(5.58\pm 0.22)\times 10^{-2}$&$1.06$&$0.926$&$7.36\times 10^{-2}$\\
$B^+\to \bar{D}^{*0} \tau^+ \nu_\tau$&&&$(1.88\pm 0.20)\times 10^{-2}$&&&$1.58\times 10^{-2}$\\
\hline
$B_s^0\to D_s^{*-}\mu^+\nu_\mu$&$1.64\pm 0.278$~\cite{Harrison:2021tol}&$0.958\pm 0.146$~\cite{Harrison:2021tol}&$(5.3\pm 0.5)\times 10^{-2}$&$1.10$&$0.959$&$5.75\times 10^{-2}$\\
\hline
\end{tabular}
\caption{Decays $P\to V\ell\bar{\nu}_\ell$ or their charge conjugates. The data for comparison are taken from PDG~\cite{ParticleDataGroup:2022pth} if not indicated otherwise.\label{table_semileptonic_vector}}
\end{table}
These tables also contain some parameters of form factors: $F_+(0)$ for final-state pseudoscalar mesons~\eqref{form_factors_pseudoscalar}, and
\begin{equation*}
r_2=\frac{A_2(0)}{A_1(0)},\quad r_V=\frac{V(0)}{A_1(0)}
\end{equation*}
for vector meson in the final state~\eqref{form_factors_vector}.

The isospin symmetry $m_u=m_d$ in the present framework is exact, so the following relations for calculated form factors of semileptonic decays hold:
\begin{align*}
F_\pm^{D^0\to \pi^-}&=\sqrt{2}F_\pm^{D^+\to\pi^0},&
F_i^{D^0\to \rho^-}&=\sqrt{2}F_i^{D^+\to\rho^0}=\sqrt{2}F_i^{D^+\to\omega},\\
F_\pm^{D^0\to K^-}&=F_\pm^{D^+\to\bar{K}^0},&
F_i^{D^0\to K^{*-}}&=F_i^{D^+\to\bar{K}^{*0}},\\
F_\pm^{B^0\to \pi^-}&=\sqrt{2}F_\pm^{B^+\to \pi^0},&
F_i^{B^0\to \rho^-}&=\sqrt{2}F_i^{B^+\to\rho^0}=\sqrt{2}F_i^{B^+\to\omega},\\
F_\pm^{B^0\to D^-}&=F_\pm^{B^+\to D^0},&
F_i^{B^0\to D^{*-}}&=F_i^{B^+\to D^{*0}},
\end{align*}
where $F_i=A_0,A_1,A_2,V$.
For the purposes of the present paper it also suffices to neglect $CP$-violation, so additionally
\begin{align*}
F_\pm^{K^-\to\pi^0}=F_\pm^{K_L\to\pi^+}=F_\pm^{K_L\to\pi^-}.
\end{align*}

The vertex~\eqref{qmvert}, and consequently~\eqref{qmvert_W}, results from the expansion of a bilocal quark current in terms of basis functions around its ``center of mass''~\cite{Efimov:1995uz}. The vertex with $W$ boson~\eqref{qmvert_W} changes quark flavor and hence mass, and this might lead to exponential growth of individual terms in the sum over radial number $n$ in formulas~\eqref{hadronic_part_semileptonic_pseudoscalar} and~\eqref{hadronic_part_semileptonic_vector}, which is typical of nonlocal theories. The sums over $n$ then appear as a result of large number cancellation which makes them difficult to evaluate numerically. Among the considered semileptonic decays, this happens with the decays of $D_s$ into $K,K^*$, $B$ into $\pi,\rho,\omega$, and $B_s$ into $K$. The impulse approximation is unaffected by this numerical instability, while additional ``nonlocal'' contributions are neglected for these decays as they are expected to be smaller. The latter is mentioned in Ref.~\cite{Faessler:2008ix} and can be observed in semileptonic decays where there is no numerical instability.

One notices that the branching ratios of semileptonic decays with light final-state vector mesons such as $D^0\to \rho^- e^+ \nu_e$ (see Table~\ref{table_semileptonic_vector}) are in poor agreement with experimental data. 
However, the experimental values given in Table~\ref{table_semileptonic_vector} are not measured directly, but rather extracted from semileptonic decays with a couple of final-state pseudoscalar mesons
\begin{equation}
\label{decay_PPPlnu}
P\to P'P''\ell\nu_\ell.
\end{equation}
In doing so, these decays are considered to take place via resonances
and the Breit-Wigner function is employed for the description of their shape (see e.g. Ref.~\cite{CLEO:2011ab}). In the Domain Model, the meson propagator~\eqref{meson_propagator} resembles the free particle propagator only in the vicinity of the meson pole. If one includes finite width in propagator~\eqref{meson_pf}, it is expected to be adequately approximated by Breit-Wigner form only for narrow resonances. Assumingly, this is why decay width for $B_{(s)}^0\to D_{(s)}^{*-}\mu^+\nu_\mu$ is in much better agreement with experiment (see Table~\ref{table_semileptonic_vector}) because the widths of $D^{*}$ and $D_{s}^{*}$ are relatively small. Another source of uncertainty is non-resonant contribution to the decay~\eqref{decay_PPPlnu} which would interfere with the resonant one.
Overall, the comparison of model result with experimental data for four-body decay~\eqref{decay_PPPlnu} would be more conclusive, but this requires evaluation of both resonant and non-resonant contributions, which is beyond the scope of the present paper. The ratios $r_2,r_V$ are not sensitive to the total decay width.

In order to conveniently represent the calculated form factors, they are fitted with the double-pole parametrization
\begin{equation}
\label{double_pole_parametrization}
F_i=F_i(0)\left[1-a\frac{q^2}{M_P^2}+b\left(\frac{q^2}{M_P^2}\right)^2\right]^{-1}
\end{equation}
in the physical region of $q^2$. The results of the fits are given in Tables~\ref{table_form_factors_pseudoscalar} and~\ref{table_form_factors_vector}, and the fitting error is negligible for the present framework. The Tables~\ref{table_form_factors_pseudoscalar} and~\ref{table_form_factors_vector} allow to compare the contribution of all diagrams in Fig.~\ref{diagrams_semileptonic_decay} with the impulse approximation where the former is accessible.
\begin{table}
\begin{tabular}{|cc|ccc|ccc|}
\hline
&&\multicolumn{3}{c|}{$F_-$}&\multicolumn{3}{c|}{$F_-$}\\
\cline{3-8}
&&$F_+(0)$&$a$&$b$&$F_-(0)$&$a$&$b$\\
\hline
\multirow[c]{2}{*}{$K^-\to \pi^0 \ell \bar{\nu}_\ell$}&full&
$0.699$&$0.279$&$0.00537$&$-0.0964$&$0.210$&$0.00703$\\
&imp. approx.&
$0.697$&$0.280$&$0.00606$&$-0.0876$&$0.224$&$0.00962$\\
\hline
\multirow[c]{2}{*}{$D^0\to \pi^- \bar{\ell} \nu_\ell$}&full&
$0.745$&$0.657$&$-0.0364$&$-0.375$&$0.813$&$0.0830$\\
&imp. approx.&
$0.648$&$0.772$&$0.0490$&$-0.382$&$0.790$&$0.0641$\\
\hline
\multirow[c]{2}{*}{$D^0\to K^- \bar{\ell} \nu_\ell$}&full&
$0.813$&$0.614$&$0.0137$&$-0.388$&$0.642$&$0.0282$\\
&imp. approx.&
$0.803$&$0.624$&$0.0205$&$-0.386$&$0.644$&$0.0287$\\
\hline
$D_s\to K^0 \bar{\ell} \nu_\ell$&imp. approx.&
$0.611$&$1.02$&$0.185$&$-0.388$&$1.05$&$0.197$\\
\hline
$B^0\to \pi^- \bar{\ell} \nu_\ell$&imp. approx.&
$0.348$&$1.08$&$0.170$&$-0.285$&$1.09$&$0.181$\\
\hline
$B_s^0\to K^- \bar{\ell} \nu_\ell$&imp. approx.&
$0.270$&$1.39$&$0.434$&$-0.230$&$1.40$&$0.434$\\
\hline
\multirow[c]{2}{*}{$B^0\to D^- \bar{\ell} \nu_\ell$}&full&
$0.839$&$0.629$&$0.0227$&$-0.382$&$0.635$&$0.0237$\\
&imp. approx.&
$0.840$&$0.629$&$0.0231$&$-0.382$&$0.635$&$0.0233$\\
\hline
\multirow[c]{2}{*}{$B_s^0\to D_s^- \bar{\ell} \nu_\ell$}&full&
$0.801$&$0.807$&$0.115$&$-0.360$&$0.840$&$0.142$\\
&imp. approx.&
$0.785$&$0.828$&$0.135$&$-0.363$&$0.832$&$0.135$\\
\hline
\end{tabular}
\caption{Parameters of double-pole parametrization~\eqref{double_pole_parametrization} fitted to form factors of $P\to P'\ell\bar{\nu}_\ell$ extracted from the amplitude~\eqref{hadronic_part_semileptonic_pseudoscalar}. Rows with label ``full'' include all terms in Eq.~\eqref{hadronic_part_semileptonic_pseudoscalar}, while in ``imp. approx'' only the term corresponding to the impulse approximation is retained.\label{table_form_factors_pseudoscalar}}
\end{table}

\begin{table}
\begin{tabular}{|cc|ccc|ccc|ccc|ccc|}
\hline
&&\multicolumn{3}{c|}{$A_0$}&\multicolumn{3}{c|}{$A_1$}&\multicolumn{3}{c|}{$A_2$}&\multicolumn{3}{c|}{$V$}\\
\cline{3-14}
&&$A_0(0)$&$a$&$b$&$A_1(0)$&$a$&$b$&$A_2(0)$&$a$&$b$&$V(0)$&$a$&$b$\\
\hline
\multirow[c]{2}{*}{$D^0\to \rho^- \bar{\ell} \nu_\ell$}&full&
$1.07$&$0.972$&$0.248$&$0.948$&$0.113$&$0.0258$&$0.777$&$0.316$&$0.196$&$1.30$&$0.785$&$0.0908$\\
&imp. approx.&
$1.14$&$0.894$&$0.168$&$0.953$&$0.110$&$0.0249$&$0.688$&$0.370$&$0.237$&$1.30$&$0.785$&$0.0908$\\
\hline
\multirow[c]{2}{*}{$D^0\to K^{*-} \bar{\ell} \nu_\ell$}&full&
$1.03$&$0.779$&$0.130$&$0.922$&$0.118$&$-0.00706$&$0.726$&$0.433$&$0.0543$&$1.14$&$0.683$&$0.0613$\\
&imp. approx.&
$1.03$&$0.762$&$0.112$&$0.921$&$0.118$&$-0.00715$&$0.715$&$0.441$&$0.0584$&$1.14$&$0.683$&$0.0613$\\
\hline
\multirow[c]{2}{*}{$D_s\to \phi \bar{\ell} \nu_\ell$}&full&
$0.856$&$1.08$&$0.352$&$0.811$&$0.349$&$-0.00735$&$0.713$&$0.569$&$0.0943$&$1.10$&$0.936$&$0.191$\\
&imp. approx.&
$0.889$&$1.00$&$0.256$&$0.809$&$0.347$&$-0.00883$&$0.633$&$0.645$&$0.154$&$1.10$&$0.936$&$0.191$\\
\hline
$D_s^+\to K^{*0} \bar{\ell} \nu_\ell$&imp. approx.&
$0.855$&$1.18$&$0.361$&$0.719$&$0.397$&$0.00367$&$0.495$&$0.580$&$0.343$&$1.05$&$1.09$&$0.267$\\
\hline
$B^0\to \rho^- \bar{\ell} \nu_\ell$&imp. approx.&
$0.670$&$1.30$&$0.373$&$0.559$&$0.294$&$0.00389$&$0.519$&$1.02$&$0.296$&$0.656$&$1.18$&$0.265$\\
\hline
\multirow[c]{2}{*}{$B^0\to D^{*-} \bar{\ell} \nu_\ell$}&full&
$0.891$&$0.769$&$0.110$&$0.843$&$0.190$&$-0.0316$&$0.780$&$0.679$&$0.0733$&$0.890$&$0.733$&$0.0817$\\
&imp. approx.&
$0.893$&$0.762$&$0.103$&$0.844$&$0.190$&$-0.0316$&$0.780$&$0.680$&$0.0738$&$0.890$&$0.733$&$0.0817$\\
\hline
\multirow[c]{2}{*}{$B_s^0\to D_s^- \bar{\ell} \nu_\ell$}&full&
$0.788$&$1.03$&$0.278$&$0.766$&$0.414$&$-0.0237$&$0.734$&$0.871$&$0.160$&$0.843$&$0.957$&$0.204$\\
&imp. approx.&
$0.805$&$0.989$&$0.235$&$0.768$&$0.412$&$-0.0246$&$0.716$&$0.894$&$0.182$&$0.843$&$0.957$&$0.204$\\
\hline
\end{tabular}
\caption{Parameters of double-pole parametrization~\eqref{double_pole_parametrization} fitted to form factors of $P\to V\ell\bar{\nu}_\ell$ extracted from the amplitude~\eqref{hadronic_part_semileptonic_vector}. Rows with label ``full'' include all terms in Eq.~\eqref{hadronic_part_semileptonic_vector}, while in ``imp. approx'' only the term corresponding to the impulse approximation is retained.\label{table_form_factors_vector}}
\end{table}

\section{Summary and outlook\label{section_summary}}

The Domain model of QCD vacuum and hadronization is a mean-field approach that allows unified description of basic low-energy properties of QCD and meson physics. In the present work, the model was applied to the leptonic and semileptonic decays of mesons. The numerical results for leptonic decay constants, semileptonic form factors and branching ratios are presented. The phenomenological description is reasonable with the exception of branching ratios $\mathcal{B}\left(P\to V\ell\bar{\nu}_\ell\right)$ for large-width mesons $V$.

When one consistently introduces the electroweak interactions into the model, the corresponding amplitudes of these decays get contributions due to vertices in Fig.~\ref{figure_nonlocal} in addition to the impulse approximation. The impulse approximation for leptonic decays of vector mesons in not even meaningful on its own. The Domain Model is obviously an effective model with limited precision, but the findings of the present work indicate that vertices in Fig.~\ref{figure_nonlocal} should be taken into account when extracting CKM matrix from the experimental data. It is plausible that the tension between determination of CKM matrix elements from inclusive and exclusive semileptonic decays, leptonic decays and semileptonic decays can be related to contributions in Fig.~\ref{figure_nonlocal}, as well as deviation from unitarity of CKM matrix (see e.g. review~\cite{HFLAV:2022esi}).

\begin{figure}
\includegraphics[scale=0.25]{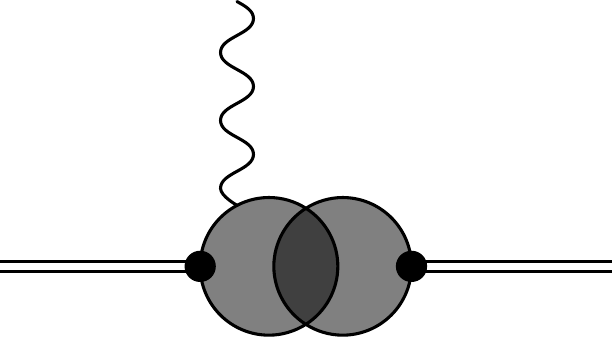}\hspace*{2em}
\includegraphics[scale=0.25]{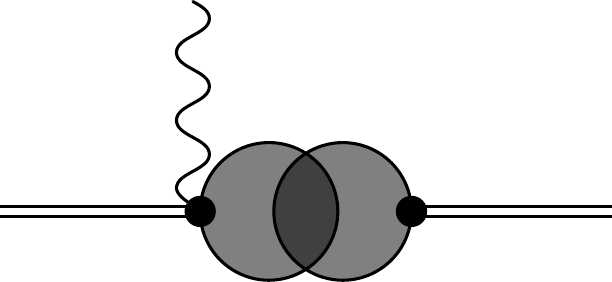}
\caption{Additional diagrams contributing to the amplitude of $P\to \eta^{(\prime)}\ell\bar{\nu}_\ell$ in the Domain model. The dark grey denotes correlation of quark loops by the vacuum field analogous to the second term in Eq.~\eqref{Gammak}.\label{figure_diagrams_eta}}
\end{figure}

The decays involving $\eta$ and $\eta'$ mesons were not considered in the present work because there are additional contributions to corresponding amplitudes. The diagrams for these contributions are shown in Fig.~\ref{figure_diagrams_eta}, their origin is analogous to the second term of Eq.~\eqref{Gammak}. These decays require a separate thorough analysis and will be considered elsewhere.

\begin{acknowledgments}
The author is grateful to Sergei Nedelko for fruitful and stimulating discussions.
\end{acknowledgments}

\appendix*
\section{Vertex function\label{appendix_vertex_function}}
\begin{equation}
\label{vertex_function}
\begin{split}
F_{nl}(s)&=s^n\int_0^1 dt t^{n+l} \exp(st)=\int_0^1 dt t^{n+l} \frac{\partial^n}{\partial t^n} \exp(st)\\
&=\left.t^{n+l}\frac{\partial^{n-1}}{\partial t^{n-1}}\exp(st)\right|_{t=0}^1-\int_0^1 dt (n+l) t^{n-1+l} \frac{\partial^{n-1}}{\partial t^{n-1}} \exp(st)=\dots\\
&=(-1)^n (n+l)! \left[\sum_{m=1}^n(-1)^m \frac{1}{(m+l)!}s^{m-1}\exp(s)+\frac{1}{l!}\int_0^1 dt t^l \exp(st) \right].
\end{split}
\end{equation}
The terms $s^n$ containing kinematic variables make formulas significantly larger. In order to avoid this difficulty, formula~\eqref{vertex_function} for vertex function $F_{nl}$ can be transformed further with the help of the identity
\begin{equation}
\label{power_identity}
s^n=\frac{n!}{2\pi i}\oint_\Gamma \frac{dz}{z^{n+1}}\exp sz,\quad n=0,1,2,\dots
\end{equation}
where the closed contour $\Gamma$ encircles zero. After substituting Eq.~\eqref{power_identity} into Eq.~\eqref{vertex_function} and using several identities one finds
\begin{multline*}
F_{nl}=(-1)^n (n+l)! \int_0^1 dt\left[\vphantom{\frac11}\exp(s(r\exp(i2\pi t)+1))\right.\\
\left.\times\left\{\sum_{m=2}^n \frac{(-1)^m}{(m+l)!}\frac{(m-1)!}{r^{m-1}}\exp(i2\pi (1-m)t)\right\}-\frac{\exp s}{(l+1)!}+\frac{t^l}{l!} \exp(st) \right]\\
=(-1)^n (n+l)! \int_0^1 dt\left[\vphantom{\frac11}\frac{t^l}{l!} \exp(st)-\frac{\exp s}{(l+1)!} +\sum_{m=2}^n \frac{(-1)^m}{(m+l)!}\frac{(m-1)!}{r^{m-1}}\frac{-i}{m-1}\times\right.\\
\left. \times\sum_{j=0}^{m-2}\left\{\exp\left[sf(t/2-j,m,r)\right]-\exp\left[sf(-t/2+j,m,r)\right]\right\}\sin\pi t\right]
\end{multline*}
where
\begin{equation*}
f(t,m,r)=r\exp\left(i2\pi \frac{t}{m-1}\right)+1
\end{equation*}
and $0<r\leqslant 1$ is an arbitrary parameter.
\bibliography{references}
\end{document}